\documentstyle[pra,aps,epsfig]{revtex}
\begin{document}
\draft
\twocolumn[\hsize\textwidth\columnwidth\hsize\csname @twocolumnfalse\endcsname
\author{M. Fran\c{c}a Santos$^1$, E. Solano$^{1,2}$ and R. L. de Matos Filho$^1$}
\title{Conditional large Fock state preparation and field state reconstruction in Cavity QED}
\address{$^1$Instituto de F\'{\i}sica, Universidade Federal do Rio de Janeiro,
Caixa Postal 68528, 21945-970 Rio de Janeiro, RJ, Brazil \\
$^{2}$Secci\'{o}n F\'{\i}sica, Departamento de Ciencias,
Pontificia Universidad Cat\'{o}lica del Per\'{u}, Apartado 1761,
Lima, Peru}
\date{February 2nd, 2001}
\maketitle
\begin{abstract}
  We propose a scheme for producing large Fock states in Cavity QED via the
  implementation of a highly selective atom-field interaction. It is based on
  Raman excitation of a three-level atom  by a classical
  field and a quantized field mode. Selectivity appears when one tunes to
  resonance a specific transition inside a chosen atom-field subspace, while
  other transitions remain dispersive, as a consequence of the field dependent
  electronic energy shifts. We show that this scheme can be also employed for
  reconstructing, in a new and efficient way, the Wigner function of the
  cavity field state.
\end{abstract}

\pacs{PACS number(s): 42.50.Dv,32.80.-t,03.65.-w}
\vskip2pc] 

Harmonic oscillators  have been, from the very beginning, at the
core of the quantum theory. It was not until the invention of the
laser~\cite{laser}, however, that their most interesting
statistical properties could be tested in controlled experiments associated
with electromagnetic fields.
Since then, considerable theoretical and experimental efforts have
been devoted to the production and characterization of
nonclassical states of light, such as sub-Poissonian~\cite{subp},
squeezed~\cite{sque}, or Schr\"odinger cat states~\cite{cat,cat2}.

A great advance in the field came with the
micromaser~\cite{microm}, where two-level Rydberg atoms interact
with one mode of a high-Q cavity in an experimental realization of
the Jaynes-Cummings (JC) model~\cite{JC}. Many important
experiments have followed, expliciting the dynamical properties of
this model, such as the observation of collapse and
revivals~\cite{rev} and the discrete character of the Rabi
oscillations~\cite{rabi} of the atom-field doublets. The
micromaser technique has also allowed the investigation of the
most fundamental, nonclassical states of harmonic oscillators,
specially the number (Fock) states. In particular, recent papers
reported the QND measurement of the one photon Fock
state~\cite{1foton} and the preparation of up to two
photons~\cite{2fotons} in the cavity mode.

The atom-field dynamics as well as the statistical properties of the field are
observed, in Cavity QED, through the detection of the atoms, which work
either as a probe for the coupled system or as a measuring device for the
light mode state, depending on the setup. Indeed, several theoretical results have showed that appropriate settings of this system enable a complete reconstruction of the quantum state of the cavity
field~\cite{Meystre,LG}. However, despite the great advances in the latest years, the implementation of
these proposals, as well as the generation and characterization of large Fock
states remain as challenging experimental problems in this field. In these
cases, the characteristics of the JC model require the use of improved
apparatus, such as cavities with higher quality factor and enhanced ways to
control and manipulate the atoms.

Alternatively, different atom-field couplings could be used. In particular,
selective interactions~\cite{kike} present new ways to induce transitions in this system. Unlike the JC model, where the same resonant or dispersive regime applies for all initial atom-field states, selective interactions separate these states in subspaces with distinct coupling regimes. This property represents more versatility and the possibility to implement new classes of experiments even under current technological conditions. 

In this paper, we show that it is possible to implement a selective
interaction between three level atoms, a classical field and a quantized
cavity mode. We study the feasibility of our scheme based on available
experimental parameters and propose, as a first application, the
preparation of large Fock states in the quantized mode. As a
second relevant application, we show that selectivity in this
system allows for the measurement of the probability distribution
$P(n)$ of the quantized field state in the Fock state basis
$|n\rangle$.  Finally, we show that it is possible, as a natural
consequence, to propose an  efficient reconstruction method of the
Wigner function~\cite{Wigner} of the cavity field and, therefore,
of its complete quantum state~\cite{Cahill}.

\vspace*{-1.3cm}
\begin{figure}[htb]
\begin{center}
\hspace*{1cm}\centerline{\leavevmode \epsfxsize=15cm\epsfbox{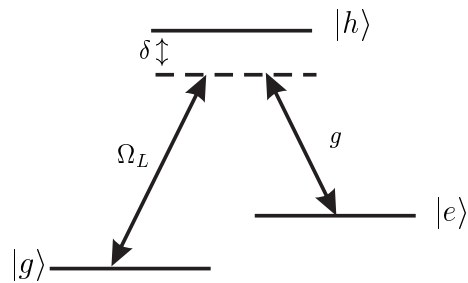}}
\vspace*{-14.5cm}
\caption{Scheme for the Raman excitation  of the three-level atom.} 
\label{fig1}
\end{center}
\end{figure}
Our proposal relies on the Raman excitation of a three level atom by a
classical field of frequency $\omega _{L}$ and a quantized cavity mode of
frequency $\omega _{0}$, in a lambda configuration (see Fig. 1). The classical
field drives dispersively the transition from level $\left| g\right\rangle$ to
level $\left| h\right\rangle$, with coupling constant $\Omega_{L}$ and
detuning $\delta =\omega _{hg}-\omega_{L}\gg |\Omega _{L}|$.  The cavity mode
couples level $\left| e\right\rangle $ to level $\left| h\right\rangle $, with
coupling constant $g$ and the same detuning $\delta =\omega _{he}\!-\!\omega
_{0}\gg |g|$. In the interaction picture, the interaction Hamiltonian in the
RWA approximation is given by
\begin{equation}
\widehat{H}_{\rm int}=\hbar \Omega _{L}\widehat{\sigma }_{hg}e^{-i\delta
  t}+\hbar g\widehat{\sigma}_{he}\widehat{a}e^{-i\delta t}+{\rm H.c.},
\label{Hini2}
\end{equation}
where $\widehat{\sigma}_{jm}\equiv| j \rangle\langle m |$ is an
electronic flip operator and $\hat{a}$ is the annihilation
operator of the quantized cavity mode. Since level $\left|
h\right\rangle $ is coupled dispersively with both levels $\left|
g\right\rangle $ and $\left| e\right\rangle $, it can be
adiabatically eliminated giving rise to an effective second order
anti-Jaynes-Cummings Hamiltonian
\begin{eqnarray}
\widehat{H}_{\rm eff} = & \hbar & \frac{ |\Omega_{L}| ^{2}}{\delta}
\widehat{\sigma }_{gg}+\hbar \frac{\left| g\right|
^{2}\widehat{a}^{\dagger}\widehat{a}}{\delta} \widehat{\sigma
}_{ee} \nonumber \\ & + & \hbar \frac{|g \Omega _{L}^*|}{\delta}
(\widehat{\sigma }_{eg}\widehat{a}^{\dagger }+\widehat{\sigma
}_{ge}\widehat{a}) \, ,  \label{Heff}
\end{eqnarray}
where we chose $\Omega _{L}$ in phase with $g$. The
first two terms correspond to dynamical energy shifts of levels
$\left| g\right\rangle$ and $\left| e\right\rangle $, and the last
two terms describe transitions between these levels, accompanied
by creation or annihilation of a photon in the cavity mode. Notice
that the difference of the energy shifts of level $| g \rangle$
and $| e \rangle$, which  depends explicitly on the number $n$ of
photons in the cavity mode, will determine the effective resonance
frequency of the $| g \rangle \leftrightarrow | e \rangle$
transition.

The Hamiltonian~(\ref{Heff}) is block separable in the subspaces
spanned by the states $\{ \left|g,n\right\rangle, \left|
e,n+1\right\rangle \}$ of the atom-field system. There is a
specific difference of energy shifts $\Delta^ {n}_{eg}$,
associated to each one of these subspaces, which may be
compensated by external action on either the atom (dc Stark shift)
or the cavity mode (by shifting its frequency). In this way,
transitions inside a chosen subspace may be tuned to resonance,
while other transitions remain dispersive, producing a selective
interaction in the atom-field Hilbert space. Once this frequency
adjustment is made for one specific subspace
$\{\left|g,N_{o}\right\rangle,\left| e,N_{o}+1\right\rangle\} $,
the detunings associated with the remaining subspaces ($n\neq
N_{o}$) change to
\begin{equation}
\Delta_n\equiv \Delta^ {n}_{eg} - \Delta^{N_{o}}_{eg}=\frac{|g|^2}
{\delta}(n-N_o).
\end{equation}
By controlling the ratio between $g$ and $\Omega_L$, the detunings $\Delta_n$
can be made large enough for considering the effective interaction in the
remaining subspaces as dispersive, i.e., $\Delta_n \gg
\frac{|g\,\Omega_{L}^*|}{\delta}$. In this case, if the atom enters the cavity
in state $\left| g\right\rangle$, it can only experiment a Rabi flip to level
$\left| e\right\rangle$ if the cavity has $N_{o}$ photons.

Hamiltonian~(\ref{Heff}) can be easily diagonalized. In particular, after the
frequency adjustment made in Eq.~(3), its stationary eigenstates are
given by the ground state $\left| e,0\right\rangle$ and the doublets
\begin{equation}
\left|\pm ,n\right\rangle =\frac{G_{n}\left| g,n\right\rangle
+\lambda_{\pm ,n}\left| e,n+1\right\rangle}{\sqrt{\lambda_{\pm
,n}^{2}+G_{n}^{2}}} \label{eig} ,
\end{equation}
with respective eigenvalues $0$ and $ \lambda _{\pm
,n}=\frac{\Delta _{n}}{2}\pm\Omega_n$. In Eq.~(\ref{eig}),
$G_{n}=\frac{|g\,\Omega_{L}^*|}{\delta}\sqrt{n+1}$ and
$\Omega_n=\sqrt{\Delta_{n}^{2}/4+ G_{n}^{2}}$.

For arbitrary values of the detunings $\Delta_n$, if the atom enters the
cavity in state $\left| g\right\rangle $, and the field is initially in state
$\left| \Phi _{0}\right\rangle =\sum_{n}c_{n}\left| n\right\rangle $, the
state of the system evolves, after an interaction time $t$, to
\begin{eqnarray}
\left| \Psi (t)\right\rangle &=&\sum_{n}c_{n}e^{-i\frac{\Delta_{n}t}{2}}[(\cos
\Omega_n t+\frac{i\Delta _{n}}{2\Omega_n}\sin \Omega_n t)
\left| g,n\right\rangle \nonumber \\
& &-\frac{iG _{n}}{\Omega_n}\sin \Omega_n t\left|
e,n+1\right\rangle]\text {.} \label{fastate}
\end{eqnarray}
In particular, if the atom is
found in state $| e \rangle$, after it has interacted with the
light fields during a time interval $\tau =
\frac{\pi\delta}{2|g\,\Omega_{L}^*|\sqrt{N_{o}+1}}$, the correlated
state of the cavity mode is
\begin{equation}
\left| \Phi _{e}(\tau)\right\rangle =\frac{c_{N_{o}}\left|
N_{o}+1\right\rangle +\sum_{n\neq N_{o}}b_{n}\left| n\right\rangle
}{\sqrt{\left| c_{N_{o}}\right| ^{2}+\sum_{n\neq N_{o}}\left|
b_{n}\right| ^{2}}}. \label{phit1}
\end{equation}
The coefficients $b_{n}$ are given by
\begin{equation}
b_{n} =\frac{c_{n}(-i)e^{-i\frac{\Delta_{n}\tau}{2}}\sin\frac{\pi}{2}
\sqrt{\frac{q}{N_{o}+1}}}{\sqrt{q}}, \label {bn}
\end{equation}
where $q=\frac{r^{2}(n-N_{o})^{2}}{4(n+1)}+1$, and $r=\frac{
|g|}{|\Omega _{L}|}$. Note that, as $q$ increases, the
coefficients $b_{n}$ become negligible compared to the coefficient
$c_{N_{o}}$ and $\left| \Phi _{e}(\tau)\right\rangle$ tends to the
Fock state $|N_{o} + 1 \rangle$. For a chosen $N_{o}$, this
condition is satisfied if $r \gg 2 \sqrt{N_{o} + 2}$. In this
limit, the Fock state $|N_{o} + 1 \rangle$ could be produced in
the cavity field, as long as $c_{N_{o}} \neq 0$. In principle, one
could use this scheme to produce any Fock state in the quantized
mode. In practice, however, one is limited by the decay time of
the cavity, $\tau_{c}$, which must be much longer than the
interaction time, $\tau$. Typically, for Rydberg atoms interacting
with high-Q microwave cavities, $g/2\pi \sim 50 \rm kHz$
\cite{1foton}. For $N_{o}\sim 10$, $\delta/2\pi\sim 1\,\rm MHz$
and $\Omega_{L}\sim \frac{g}{30}$ ($r\sim 30$),  $\tau$ will be of
the order of $1\, \rm ms$. Interaction times of this order are
still much shorter than the decay time of the microwave cavities
with highest quality factor used nowadays (of the order of 0.3 s
\cite{Walther2}). This suggests that, in principle, our proposal
could be implemented immediately.

We may define the fidelity of generating the selected Fock state $|N_{o} + 1
\rangle$ as $F\equiv|\langle N_{o}+1 |\Phi_{e}(\tau)\rangle|^2$. $F$ is
approximately given by $1-\sum_{n}\frac{|b_{n}|^2}{|c_{N_{o}}|^2}$ and it
approaches unity when the coefficients $b_n$'s go to zero. Initial states for
which $c_{N_{o}}>c_{n}$ enhance the protocol efficiency. In this sense, good
candidates for initial cavity state are coherent states with mean number of
photons around $N_0$. Not only do they satisfy $c_{N_{o}}> c_{n}$, but they
are also easily produced in microwave cavities, by just coupling them to a
microwave generator~\cite{cat2}. In Fig. 2, we show an example for the
preparation of large Fock states in the cavity mode, after only one atom has
interacted with the fields. From an initial coherent state $| \alpha \rangle$
with $|\alpha|^2=5$, the Fock state $| 6 \rangle$ is prepared in the cavity
with a fidelity higher than $0.99$.

\vspace*{-1.7cm}
\begin{figure}[htb]
\begin{center}
\hspace*{1cm}\centerline{\leavevmode \epsfxsize=15cm\epsfbox{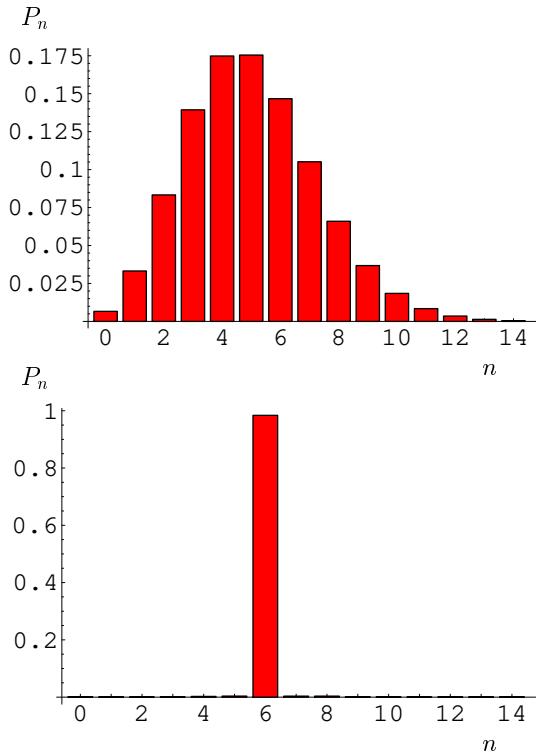}}
\vspace*{-8.5cm}
\caption{Preparation of the Fock state $| 6 \rangle$ in the cavity mode 
  by measuring the atom in its excited state after it passed through the
  cavity. The cavity field was initially in a coherent state with
  $|\alpha|^2=5$. The value of the parameter $r$ was set to $r=30$.}
\label{fig2}
\end{center}
\end{figure}

The Fock state preparation is conditioned
to finding the atom in state $| e \rangle$. From Eqs.  (\ref{fastate}),
(\ref{phit1}) and (\ref{bn}), it is easy to show that as $r$ becomes larger,
the probability $P_e$ of measuring the atom in the excited state approximates
the probability $P_{N_{o}}$ of finding $N_{o}$ photons in the initial cavity
field state. If $r$ cannot be made large enough due to experimental
limitations, the state one wants to prepare is polluted by marginal Fock
states populations. In this case, one only needs to send a second atom in
state $\left|g\right\rangle$ and set the experimental parameters to the
transition $\left|g,N_{o}+1\right\rangle \rightarrow \left|
  e,N_{o}+2\right\rangle $. The probability of finding both atoms in the
excited state becomes closer to $P_{N_{o}}$, and the field state produced will
be, with very high fidelity, state $\left|N_{o}+2\right\rangle$.

The equivalence between $P_e$ and $P_{N_{o}}$ for large $r$ suggests a very
practical and easy way to obtain the photon statistics $P_n$ of an arbitrary
state in the cavity mode. In fact, for each selected transition ${| g,N
  \rangle \leftrightarrow | e, N+1 \rangle}$, the proportion of atoms measured
in state $|e \rangle$, $P_e$, gives directly $P_{N}$, for all possible values
of $N$. Combined with the possibility of coherently displacing microwave
cavity fields, this allows one to fully reconstruct the Wigner function of the
state of the quantized mode. Since it does not rely on additional devices,
such as Ramsey interferometers, this scheme simplifies the task of field state
reconstruction, as we will discuss below.

The Wigner function of the
state $\hat{\rho}$ of a harmonic oscillator, can be written as
\begin{equation}
W(-\alpha )=\frac{2}{\pi}\sum_{n}(-1)^{n}P_n(\alpha), \label{Wigner}
\end{equation}
where $P_n(\alpha)=\left\langle n\right|\widehat{D}(\alpha
)\hat{\rho}\widehat{D}^{-1}(\alpha )\left| n\right\rangle$ is the number
distribution of state $\hat{\rho}$ displaced coherently in the phase space by
$\alpha$ \cite{Cahill,counting}. This tells us that, to obtain the Wigner
function of the cavity field on each point of the phase space, all one needs
to know is the number distribution of the field, after it has been displaced in
the phase space. The coherent displacement of the cavity field can be easily
implemented by coupling the cavity to a microwave generator. After this step,
one can use the selective scheme discussed above to measure $P_n(\alpha)$ and,
then, Eq.~(\ref{Wigner}) to calculate $W(-\alpha)$. This method is exact for
large values of $r$, and it represents an experimentally simple way to measure
the photonic statistics of the cavity field and to reconstruct its Wigner
function. For once, it does not require the preparation of atoms in a coherent
superposition of upper and lower states, as in usual schemes that rely on Ramsey interferometry. This simplifies the experimental setup and, more important, permits the use of closed higher quality cavities with characteristic lifetimes much longer than the interaction times in question. As an example, we show in Fig.~3 the efficient reconstruction of the Wigner function of the Fock state $|6\rangle$ for the realistic parameter $r=30$.  By subtracting the exact Wigner function from the reconstructed one, we also show that the errors introduced by suposing that transitions occur only inside each selected subspace ${| g,N \rangle \leftrightarrow | e, N+1 \rangle}$ (perfect selectivity) are negligibly small.

In conclusion, we have proposed a scheme to implement a selective interaction
in cavity QED, between three level atoms, a classical field and a quantized
cavity mode. It relies on the possibility of tuning to resonance a specific
transition inside a chosen atom-field subspace, while other transitions remain
dispersive, as a consequence of the field dependent electronic energy shifts.
As a first relevant application of this scheme, we have proposed a method for
generating large Fock states in the cavity mode. Aditionally, we have shown
that this scheme allows for the reconstruction of the Wigner function of an
arbitrary cavity field state, with simplified experimental setup.  Many other
application examples could be imagined, specially those exploring the
entanglement created between the atoms and different field states, as is the
case in quantum logic and quantum communication schemes.

We want to thank the hospitality of the organizers of the
Pan-American Advanced Study Insitute on "Chaos, Decoherence and
Quantum Entanglement" and S. Haroche for fruitful discussions about the experimental implementation of the proposed scheme. This work has been supported by Conselho Nacional de
Desenvolvimento Cient\'\i fico e Tecnol\'ogico (CNPq),  Funda\c
c\~ao de Amparo \`a Pesquisa do Estado do Rio de Janeiro (FAPERJ),
Funda\c c\~ao Universit\'aria Jos\'e Bonif\'acio (FUJB), and
Programa de Apoio a N\'ucleos de Excel\^encia (PRONEX).

\vspace*{-1.1cm}
\begin{figure}[htb]
\begin{center}
\hspace*{0cm}\centerline{\leavevmode \epsfxsize=15cm\epsfbox{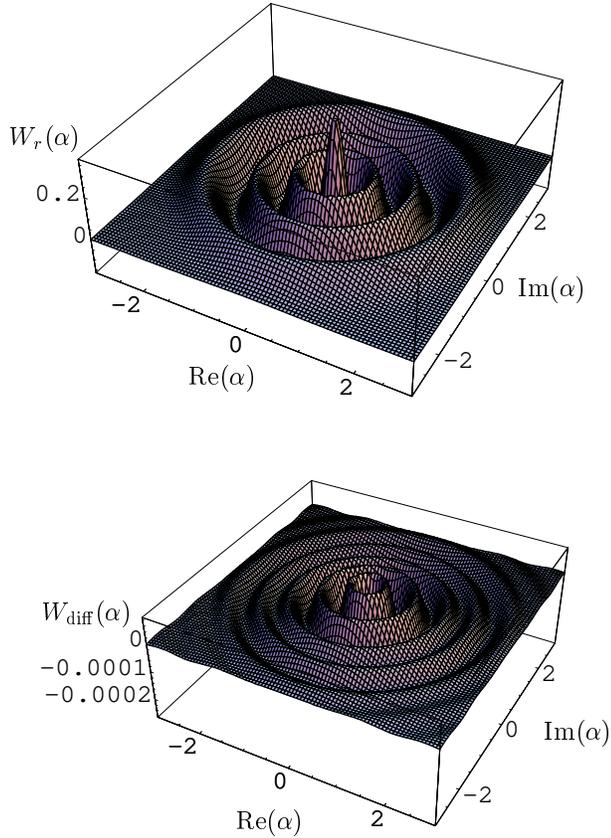}}
\vspace*{-6.5cm}
\caption{Plot of the reconstructed Wigner function of the Fock
  state $|6\rangle$ with the use of the selective scheme (above) and its
  difference to the exact one (below). The value of the parameter $r$ was set
  to $r=30$.}
\label{fig3}
\end{center}
\end{figure}

\end{document}